\documentclass[journal]{new-aiaa}
\usepackage[utf8]{inputenc}
\usepackage{textcomp}
\usepackage{wrapfig}
\usepackage{url}
\usepackage{graphicx}
\usepackage{booktabs}
\usepackage{amsmath}
\usepackage{subfig}
\usepackage[version=4]{mhchem}
\usepackage{siunitx}
\usepackage[noend]{algcompatible}
\usepackage{algorithm} 
\usepackage{longtable,tabularx}
\usepackage{threeparttable}
\algnewcommand\INPUT{\item[\textbf{Input:}]}%
\algnewcommand\OUTPUT{\item[\textbf{Output:}]}%

\title{Watercraft as Overwater Ambulance Exchange Points to Enhance Aeromedical Evacuation}

\author{Mahdi Al-Husseini,\footnote{MS Student, Aeronautics \& Astronautics Dept., Stanford University, Stanford, CA 94305.} 
Kyle H. Wray,\footnote{Visiting Scholar, Aeronautics \& Astronautics Dept., Stanford University, Stanford, CA 94305.}
Mykel J. Kochenderfer,\footnote{Associate Professor, Aeronautics \& Astronautics Dept., Stanford University, Stanford, CA 94305.}}
\affil{Stanford Intelligent Systems Lab, Stanford, CA, 94305}
\affil{25th Combat Aviation Brigade,  Wheeler Army Airfield, HI, 96857}

\begin{document}
\pagenumbering{gobble}
\maketitle
\noindent



\section*{Abstract}
\pagenumbering{gobble}
\begin{abstract}
Ambulance exchange points are preidentified sites where patients are transferred between evacuation platforms while en route to enhanced medical care. We propose a new capability for maritime medical evacuation, which involves co-opting underway watercraft as overwater ambulance exchange points to transfer patients between medical evacuation aircraft. We partner with the United States Army's 25th Combat Aviation Brigade to demonstrate the use of an Army watercraft as an overwater ambulance exchange point. A manikin is transferred between two HH-60 Medical Evacuation Black Hawk helicopters conducting hoist operations over Army Logistics Support Vessel 3, which is traveling south of Honolulu, Hawaii. The demonstration is enabled by a decision support system for dispatching aircraft, hoist stabilization technology, commercial satellite internet, military geospatial infrastructure applications, and digital medical documentation tools, the benefits of which are all discussed. Three extensions of the overwater ambulance exchange point are introduced and civilian applications are considered. 
\end{abstract}

\clearpage

\section*{Acknowledgments}
We would like to acknowledge Samuel Jones and Travis Adams at the United States Army Aeromedical Research Laboratory and Alec Finlay at Program Director MEDEVAC. These three were instrumental in providing the hoist stability hardware and training that enabled confined area SKEDCO hoist over watercraft without needing a tagline. We are grateful for our team members at the 8th Theater Sustainment Command and 545th Harbormasters, to include Jose Gomez, Teresa Kasper, and Timothy Long. They permitted us to employ their LSV-3 and LSV-7 watercraft as AXPs to execute the overwater patient transfers. We also could not have proceeded without the C/3-25 Lightning DUSTOFF aircrews who trained for several months in the lead up to execution; they include Robert Martinez, Dawn Herron, Nathan Spann, James Hill, Hunter Hardy, Eric Williams, Michael Dalleva, Andrew Briones, Antonio Gattis, Logan Davis, Phil Haake, Jacob Lipson, and Jimmy Sosa. Finally, we would like to thank 3-25 Battalion Commander Samuel Diehl, whose vision for maritime evacuation operations and expertise in all things medical evacuation greatly contributed to this paper. 


\clearpage
\pagenumbering{arabic}
\section{Introduction}
\lettrine{M}{edical} evacuation efforts in non-contiguous maritime environments like the Indo-Pacific must overcome extreme distances and various transportation constraints. The United States Army describes medical evacuation as the ``timely and effective movement of the wounded, injured, or ill to and between medical treatment facilities on dedicated and properly marked medical platforms with en route care provided by medical personnel'' \cite{medevac2019}. The Indo-Pacific region spans the Indian and Pacific Oceans and features numerous archipelagos to include the Hawaiian, Philippine, and Indonesian islands. The expedient movement and en route care of patients across these dispersed island chains and clusters requires the strategic coordination of multiple different evacuation platforms in real time. Ambulance exchange points (AXPs) are preidentified locations, typically fixed in place, where patients are exchanged from one evacuation platform to another \cite{medevac2019}. This paper considers the use of watercraft as overwater AXPs for medical evacuation aircraft transporting patients between distant islands. The underway nature of watercraft complicates their use as overwater AXPs for medical planning. Unlike traditional land-based AXPs at fixed sites, watercraft are constantly shuttling between ports, typically in support of non-medical mission requirements. However, overwater AXPs enable range-limited medical evacuation aircraft to bridge significant maritime distances to facilitate patient transfers. The positioning and selection of overwater AXPs also balances utilization across evacuation units on different islands.

\begin{figure*}[ht!]
\includegraphics[width=16.5cm]{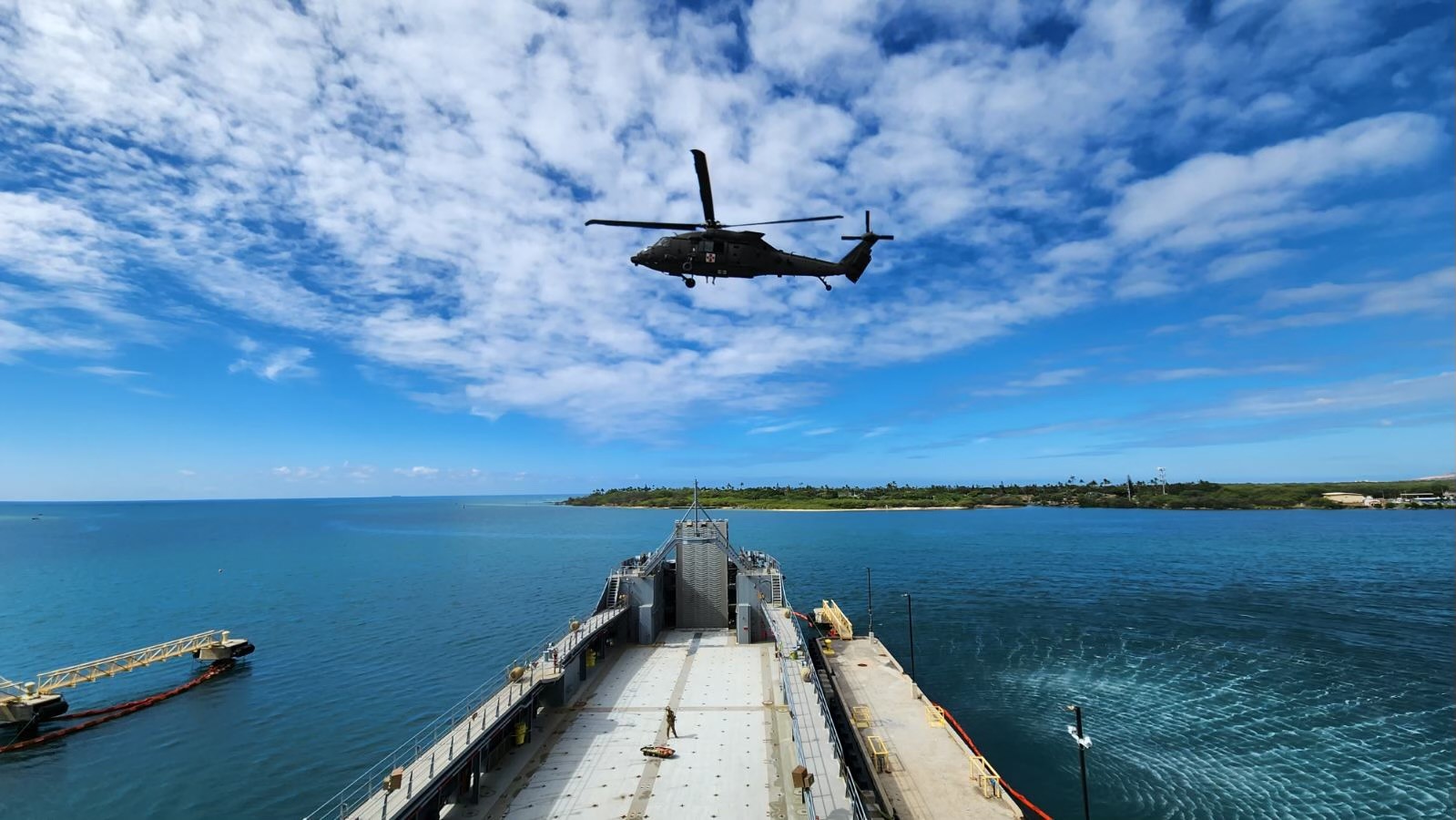}
\caption{An HH-60M Medical Evacuation Black Hawk helicopter hoists a stabilized litter patient to Army Logistics Support Vessel 7, docked in Pearl Harbor. $\copyright$ Charlie Clark 2023}
\end{figure*}
\indent
The use of aircraft to evacuate patients \textit{from} watercraft is neither novel nor remarkable. When this occurs, the watercraft is typically the point of injury and the patient requires evacuation from ship to shore. The use of aircraft to evacuate patients \textit{to} watercraft is less common. Shore to ship operations in the military are limited to casualty receiving and treatment ships which function as a role two afloat \cite{lin2009surgery}, large hospital ships such as the United States Navy's Comfort and Mercy which both function as a role three afloat \cite{drayton2013nursing}, or general area support for an amphibious force \cite{medevac2019}. The role two is a mobile military treatment facility that services a brigade-sized element or larger and provides basic primary care along with some combination of optometry, behavioral health, dental, laboratory, X-ray, and surgery. The role three expands on the capabilities provided at the role two and may contain hundreds of beds at the expense of mobility \cite{cubano2018emergency}. There is however little precedent for what we call the shore to ship to shore patient transfer. Shore to ship to shore transfers evacuate patients to watercraft, not for care but for collection en route to a more capable hospital, at scale. En route care may be provided during both the shore to ship and ship to shore legs, in which case medical evacuation is provided, or during one or none of those legs, in which case there is limited continuity of care and therefore casualty evacuation is provided. As will be demonstrated, the employment of watercraft as overwater AXPs has remarkable advantages, but requires the careful coordination of platforms and enhanced battlefield situational awareness to properly execute. 

The military community has expressed renewed interest in maritime operations for health service and support operations, spurred by a focus on joint-service integration and the Indo-Pacific theater \cite{lundy2018meeting}. The academic literature, while sparse on multi-modal maritime evacuation modeling, introduces adjacent efforts. \citeauthor{marler2022decision} applies an online optimization algorithm using boolean logic to efficiently route a single medical watercraft responsible for evacuating patients, in response to a random adversary \cite{marler2022decision}. \citeauthor{marler2022decision} identifies key challenges with maritime evacuation, to include partial observability, unknown demand location and quantities, and unpredictable environmental dynamics associated with enemy threat and poor weather. \citeauthor{carey2002future} discuss design considerations and constraints for future naval medical evacuation platforms, to include, presciently, arguing for what would eventually become the expeditionary fast transport medical ship. Special attention is given to compatibility with existing and proposed medical evacuation aircraft \cite{carey2002future}. \citeauthor{ward2008optimizing} introduces a high-fidelity optimization model for routing helicopters, watercraft, and ground vehicles between hospital ships and ashore locations to enhance personnel and patient movement during humanitarian assistance and disaster relief efforts \cite{ward2008optimizing} .

This paper's contributions are twofold. We first demonstrate the overwater AXP in a shore to ship to shore patient transfer in the Hawaiian Islands using two HH-60M Black Hawk helicopters, an Army Logistics Support Vessel (LSV), and several enabling technologies. Second, we consider novel extensions of the overwater AXP to include maritime evacuation chains, maritime transfer opportunity zones, and dedicated overwater AXPs, and discuss relevant civilian applications. 

\section{Exercise}

\subsection{MEDEVAC Projects Week Overview}

The integration of overwater AXPs, stabilized hoist operations, and transport telemedicine over secure satellite communication enables a substantial expansion in coverage of an otherwise constrained medical evacuation network in non-contiguous maritime environments. 
Artificial intelligence research at Stanford University and the Army Research Lab, partnered through a cooperative research and development agreement, suggests that using watercraft as overwater AXPs can significantly improve patient outcomes. The inaugural 2023 MEDEVAC Projects Week brought together elements of the 25th Combat Aviation Brigade, Army Futures Command, 8th Theater Sustainment Command, and the 18th Theater Medical Command to demonstrate, for the first time, the watercraft as overwater AXP. In the exercise's culminating event, a first HH-60M MEDEVAC Black Hawk was dispatched from Wheeler Army Airfield, Hawaii to an open field in a training area roughly four miles due east. A manikin packaged in a SKEDCO litter was received from a small ground force contingent and hoisted into the first HH-60M. The first HH-60M proceeded south to Mamala Bay to deliver the manikin to LSV-3, which was actively underway to the east at five knots. A second HH-60M was dispatched from Wheeler Army Airfield so as to arrive at LSV-3 shortly after manikin delivery. On arrival, the second HH-60M began to hoist the manikin up from LSV-3. After securing the manikin in the cabin, the second HH-60M delivered the manikin to Tripler Army Hospital. Hoist stabilization systems on both aircraft were used for all hoist operations involving the manikin. Transport telemedicine tablets allowed medical data on the manikin to be transferred between medics located on both aircraft, the watercraft, and at the hospital. Aircraft dispatch times were informed by a decision support system designed to enable overwater AXP operations, and were provided to both aircrews in an execution checklist. 

\begin{figure*}[ht!]
\includegraphics[width=16.5cm]{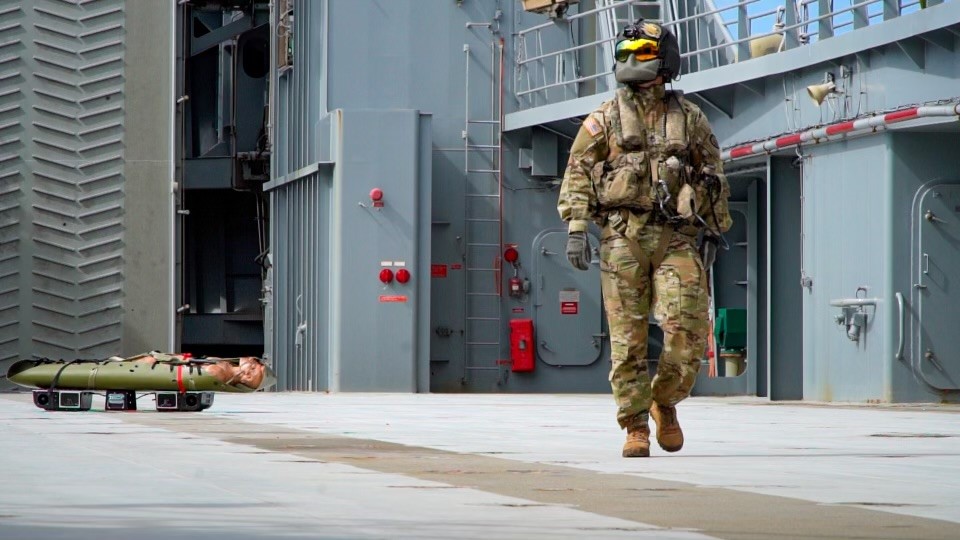}
\caption{An Army critical care flight paramedic, hoisted down from an HH-60M Medical Evacuation Black Hawk helicopter alongside his mock patient in a SKEDCO, walks the main deck of Logistics Support Vessel 7. $\copyright$ Charlie Clark 2023}
\end{figure*}

\begin{figure*}[ht!]
\includegraphics[width=16.5cm]{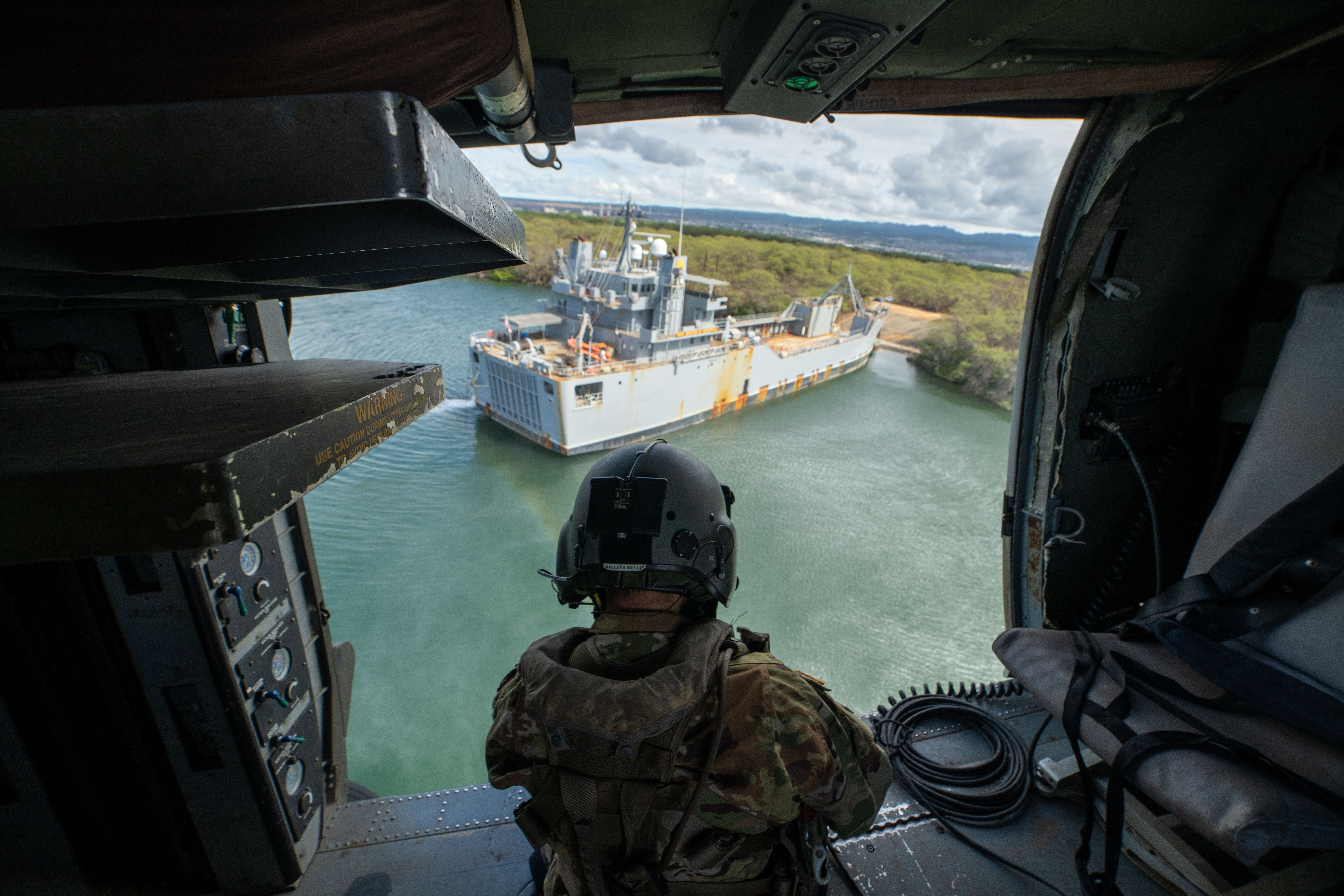}
\caption{An Army crew chief spots Logistics Support Vessel 3 from the cabin of an HH-60M Medical Evacuation Black Hawk helicopter. Logistics Support Vessel 3 would later that day depart Pearl Harbor for the open ocean. It would there serve as an ambulance exchange point for a mock patient transfer demonstration between two HH-60M helicopters. $\copyright$ Charlie Clark 2023}
\end{figure*}

\subsection{Enabling Technologies}
We introduced four enabling technologies to meet key assumptions and ensure safety during operations. Although not explicitly required to conduct patient transfers using an overwater AXP, their collective use minimized delays, reduced system uncertainty, and improved overall patient care.

\begin{figure}%
    \centering
    \subfloat[\centering ][$\copyright$ Jimmy Sosa 2023]{{\includegraphics[width=7.5cm]{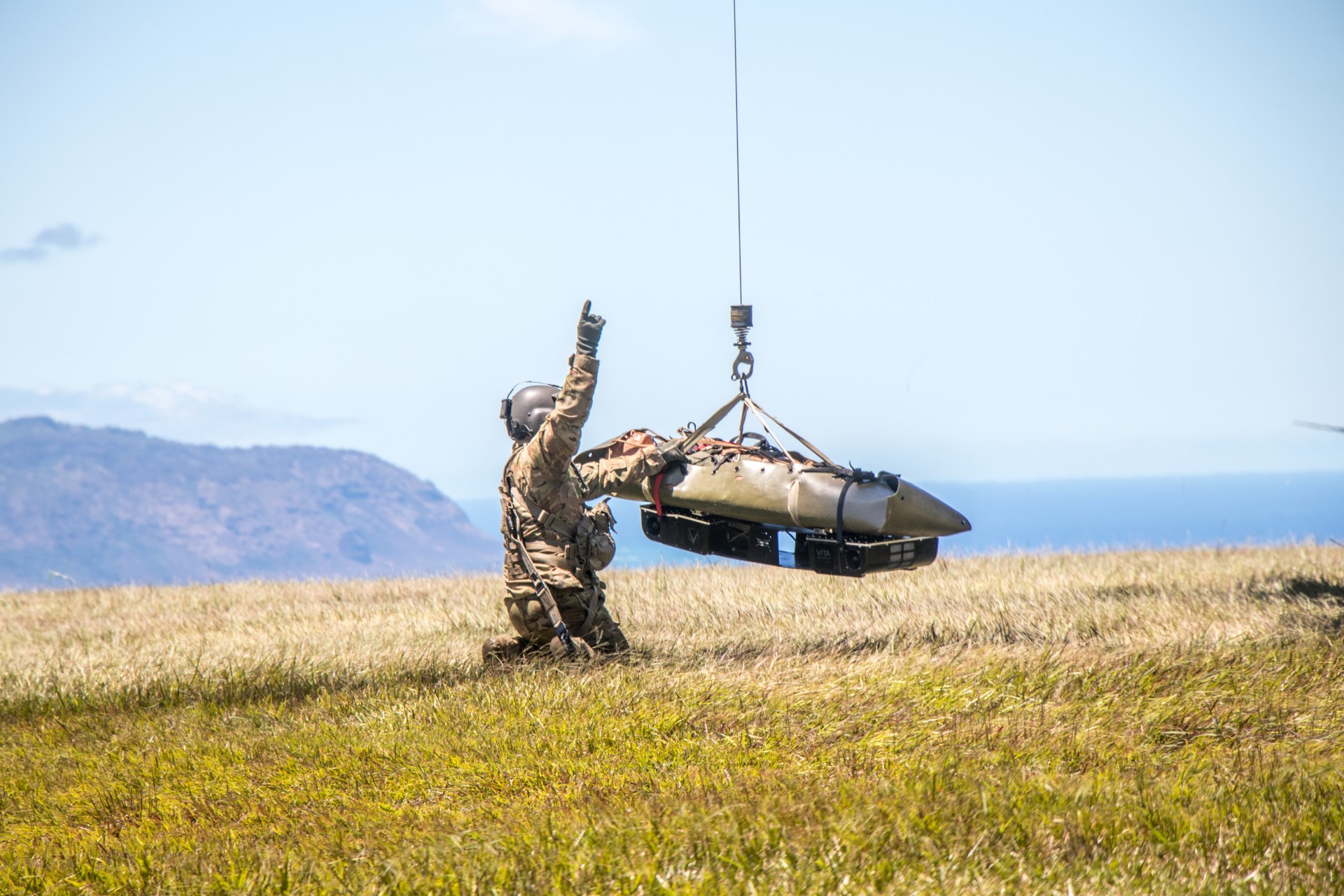} }}%
    \qquad
    \subfloat[\centering ][$\copyright$ Jimmy Sosa 2023]{{\includegraphics[width=7.5cm]{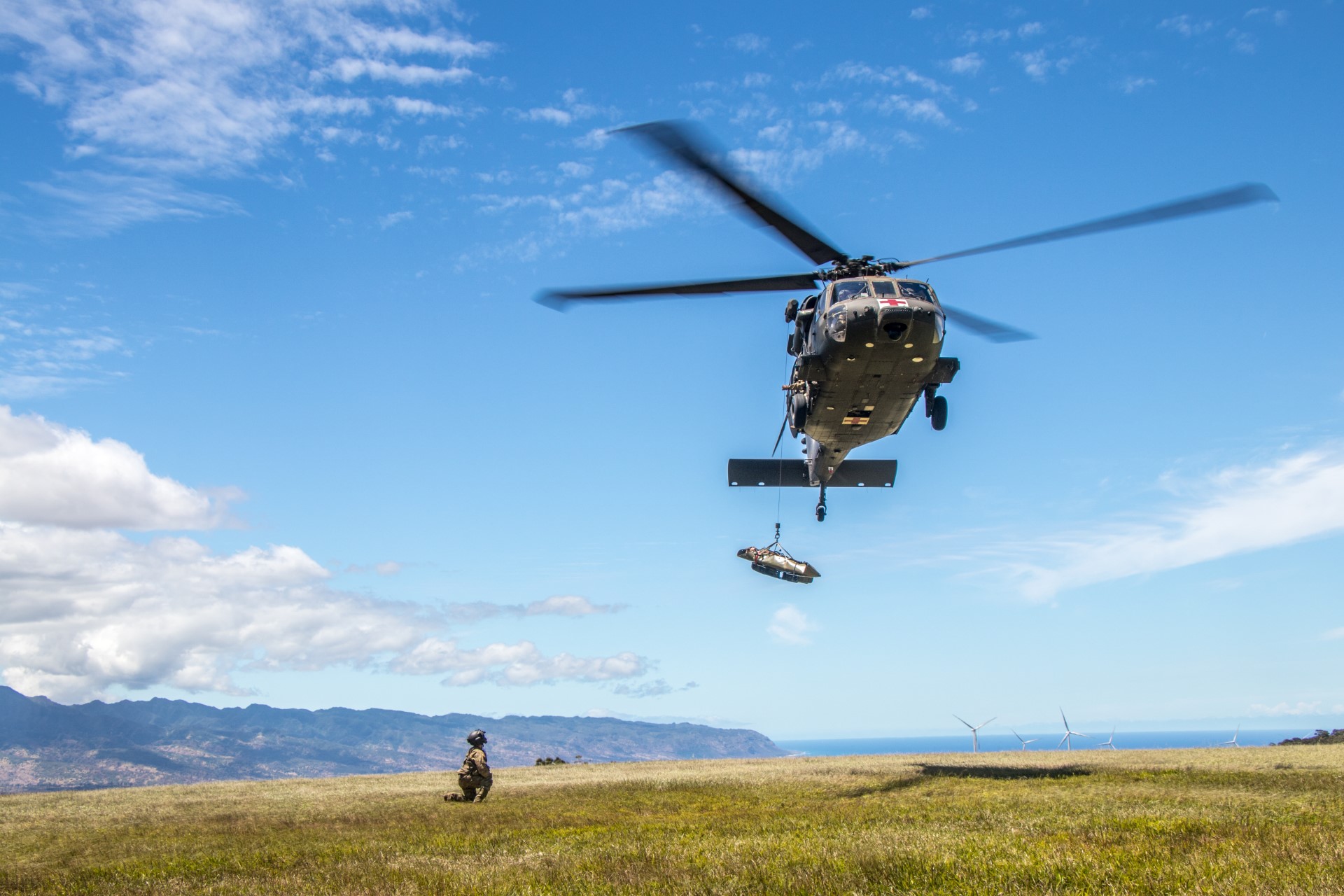} }}%
    \caption{The Load Stability System stabilizes a packaged manikin being evacuated from the point of injury. Hoist stabilization allows aircrews to hoist litter patient without using a tagline. This in turn enables hoist operations for litter patients in small confined areas such as Army watercraft.}%
\end{figure}

\subsubsection{Decision Support System to Coordinate Evacuation Platforms in a Maritime Environment}

Coordinating patient movement in time and space using a large evacuation system with multiple evacuation platforms is a difficult task for any medical planner. Evacuation platform selection and dispatching decision support systems can expedite patient movement and provide much-needed evacuation reach and flexibility. We introduce a tailored semi-Markov decision process and maritime environment model, and apply an online, stochastic sampling algorithm to determine when to dispatch evacuation aircraft to select overwater AXPs such that patient transfer times are minimized. The maritime environment model includes the Hawaiian Islands of Oahu and Kauai, two aeromedical evacuation platoons, three moving watercraft, and eight military treatment facilities. As the MEDEVAC Projects Week demonstration consisted of only two evacuation aircraft, one watercraft, and one military treatment facility, our model proved more than capable of determining the appropriate dispatch times for both aircraft. Delay between manikin drop-off and pickup at the underway LSV-3 was less than three minutes. 

\subsubsection{Hoist Stabilization Technology}
Hoist stabilization technology permits patient drop-off and pickup over the small decks of many Army watercraft. The Load Stability System uses ducted fans to automatically minimize hoist cable oscillation, spin, and sway without requiring a tagline for litter patients \cite{sikora2021bidirectional}. Taglines are suitable for large open areas, but are generally not feasible in confined spaces. The Army Transportation Corps has a fleet of approximately 132 watercraft \cite{aws2024}, none of which possess a helipad for aircraft landing and takeoff. This necessitates the use hoist operations to facilitate patient transfers over Army watercraft such as LSVs and LCUs. The majority of Navy ships, to include the EPF, have at least a single helipad. Participating aircraft will always land to deliver and collect patients when aviation-capable ships are utilized as AXPs. Due to safety and time considerations, hoist operations over watercraft will only be conducted when required. Other hoist stabilization technologies involve airfoils and sails \cite{cicolani2002modeling} or reaction-wheels inducing a counter-torque on the load \cite{al2022device}, \cite{forden2020spin} and may be applied to similar effect. Both aircrews participating in the demonstration successfully used the Load Stability System in tandem with dynamic hoist operations to efficiently and expediently lower and raise the manikin to and from the deck of the moving LSV with no recognizable instability on the hoist line. 

\begin{figure}%
    \centering
    \subfloat[\centering ][$\copyright$ Charlie Clark 2023]{{\includegraphics[width=7.5cm]{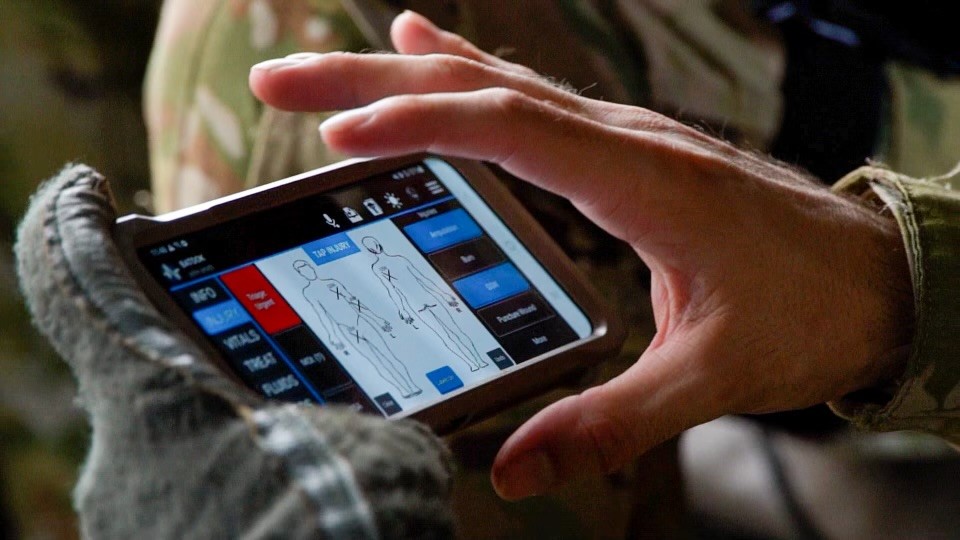} }}%
    \qquad
    \subfloat[\centering ][$\copyright$ Charlie Clark 2023]{{\includegraphics[width=7.5cm]{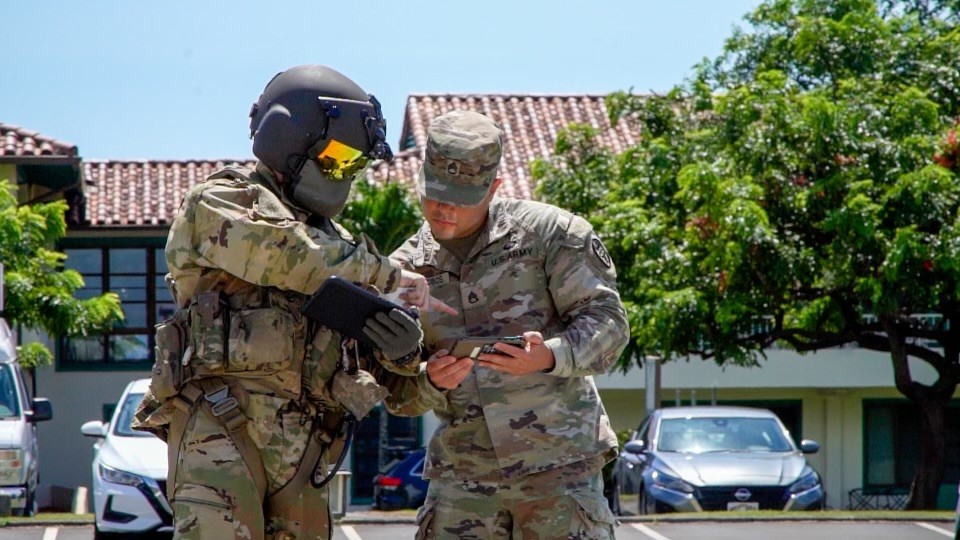} }}%
    \caption{Battlefield Assisted Trauma Distribution Kit systems were provided to the five units participating in the patient transfer process: with the ground medic at the point of injury, with the flight paramedic on the first HH-60M, with the combat medic on LSV-3, with the flight paramedic on the second HH-60M, and with the medical provider at Tripler Army Hospital.}%
\end{figure}

\subsubsection{Mobile Medical Documentation}

The patient handover process is sufficiently difficult as is \cite{raeisi2019challenges}, and the use of watercraft as overwater AXPs complicates it further. We integrate a mobile medical documentation tool called the Battlefield Assisted Trauma Distributed Observation Kit (BATDOK) to minimize the documentation burden on participating ground medics and flight paramedics and also help expediently transfer medical documentation for one or more patients between medics on different platforms. BATDOK was developed by the Air Force Research Laboratory's 711 Human Performance Wing to capture, update, and transfer patient data digitally \cite{smith2013medical}. A total of five BATDOK systems were provided to participating medical personnel; one flight paramedic on each of two aircraft, the ground medic at the inland point of injury, the medic assigned to the watercraft AXP, and the medical provider receiving the manikin at the hospital. A plastic near field communication (NFC) card is assigned to and kept with the manikin. The NFC card is repeatedly scanned to both upload and download medical information to and from each of the five BATDOK systems during patient exchange. 

\begin{figure}%
    \centering
    \subfloat[\centering ][$\copyright$ Mahdi Al-Husseini 2023]{{\includegraphics[width=5.5cm]{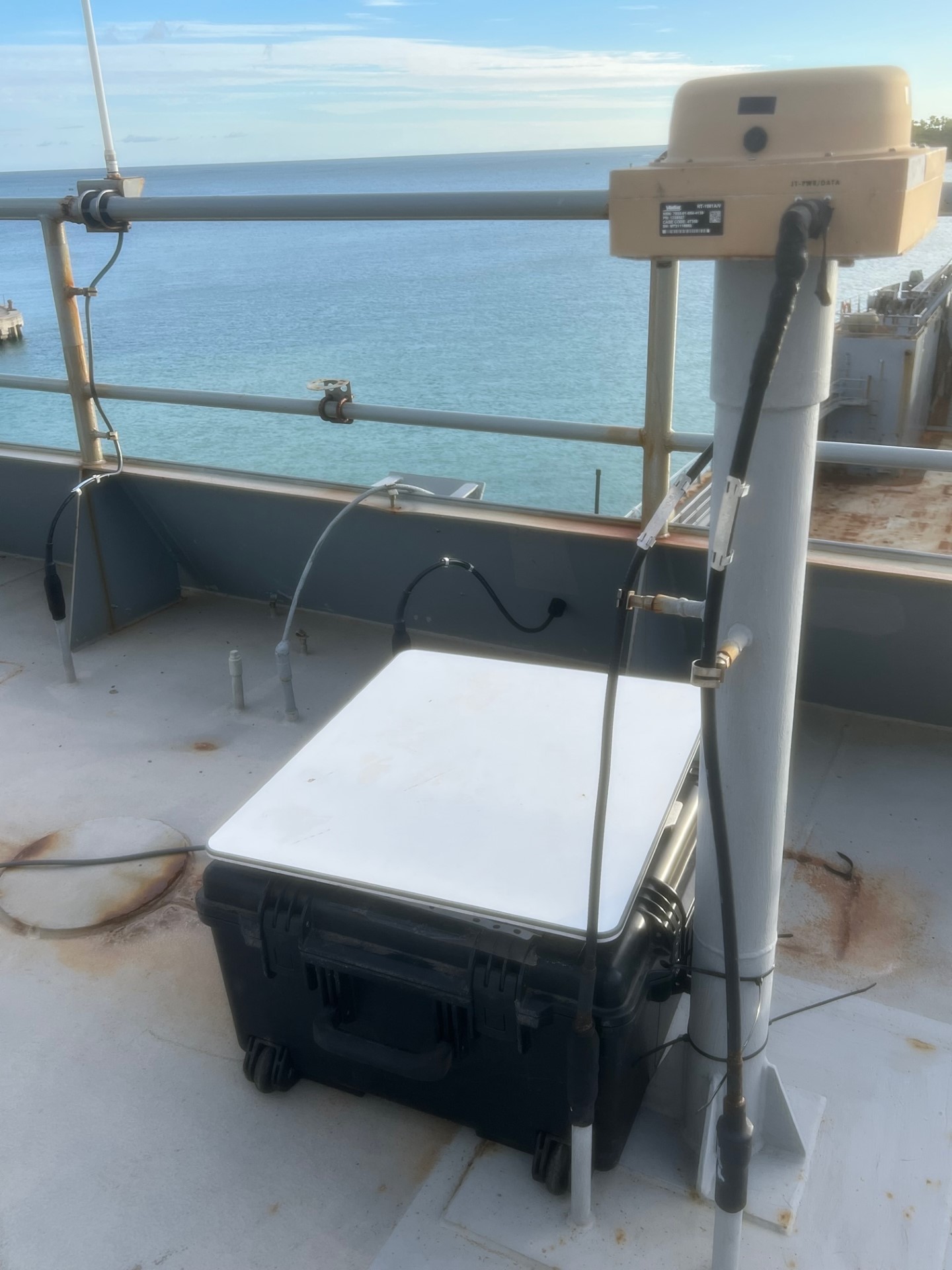} }}%
    \qquad
    \subfloat[\centering ][$\copyright$ Mahdi Al-Husseini 2023]{{\includegraphics[width=9.8cm]{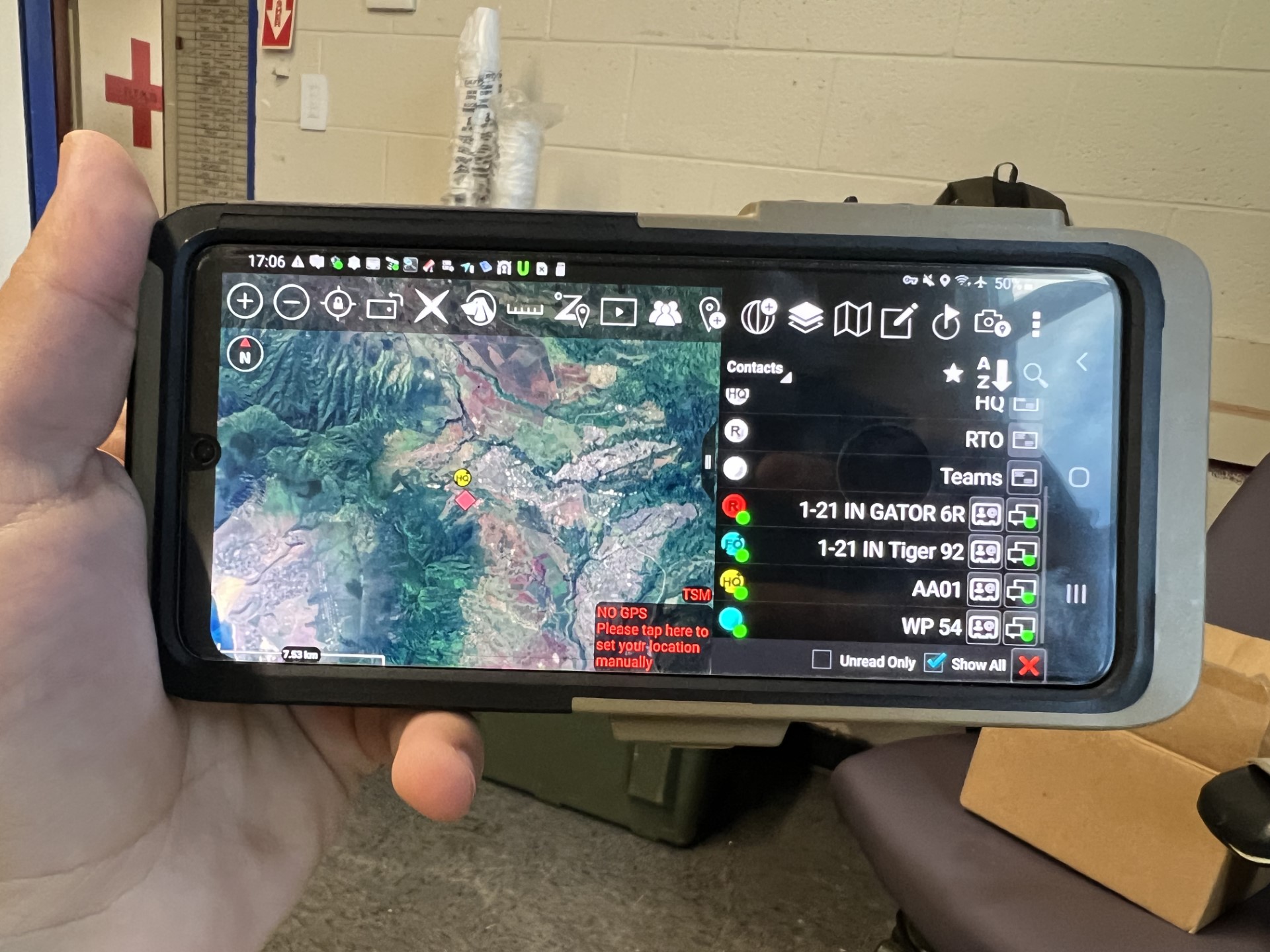} }}%
    \caption{Starlink commercial satellite internet was used in conjunction with the Android Tactical Assault Kit to transmit the evacuation request and facilitate command and control between two different command posts; one on the moving watercraft and the other at the airfield. The location of the participating watercraft and aircraft as identified on the Android Tactical Assault Kit inform the optimized dispatching model.}%
\end{figure}

\subsubsection{Satellite Internet and Tactical Assault Kit}

Full observability is an important condition for optimizing AXP selection. Watercraft and aircraft positions should be identified in real-time with reasonable certainty to select the optimal AXP in a set of AXPs for a given evacuation request. We provide two Android Tactical Awareness Kits (ATAK) to the MEDEVAC command post on Wheeler Army Airfield and the watercraft command post on LSV-3. ATAK is an Android-based geospatial infrastructure application, also developed by the Air Force Research Laboratory, that visualizes the location of participating units and assets and permits communication between them \cite{ATAK}. The ATAK-provided locations of participating aircraft and watercraft helped inform our model state. ATAK requires network connectivity to function as intended. We accomplished this using Starlink satellite communication dishes and routers installed on the bridge of LSV-3 and on Wheeler Army Airfield. ATAK was also used to transmit the evacuation request from the ground force to the MEDEVAC command post, and to communicate the exercise checklist pro-words between both command posts throughout the patient transfer for enhanced command and control. 

\section{Discussion}

\subsection{Limitations}

The use of overwater AXPs for medical evacuation is limited by jamming, watercraft with rudimentary medical capabilities, watercraft with changing routes, and participating aircraft cabin sizes. As demonstrated in the Russio-Ukrainian war, any large scale future conflict will involve degraded and denied communications via jamming \cite{mccrory2020russian}. Evacuation models must be developed that can operate near-optimally with partial observability for extended periods of time. We leave robustness, a model's ability to preform well in the face of significant environmental uncertainty or perturbation, as a critical consideration for later work. The patient handover between aircraft at the overwater AXP is near-immediate in ideal circumstances. Should delays occur however, the patient at the AXP will require an appropriate level of medical supervision to maintain continuous medical care throughout all phases of transport. Many smaller watercraft platforms have only rudimentary medical capabilities, and possess a single combat medic specialist. We assume to know the routes taken by participating watercraft in the demonstration, but changes in watercraft trajectory - before and after patient collection by aircraft - can disrupt evacuation operations and significantly increase incident response times. Finally, the patient throughput of the overwater AXP is limited to the cabin sizes of participating aircraft. We expect the overwater AXP to play a critical, if specific, role in any future maritime conflict. However, overwater AXPs will need to be scaled significantly to appropriately support the substantial number of casualties expected in a large-scale combat operation \cite{remondelli2023casualty}. 

An expansive military campaign across a noncontiguous maritime environment with several heterogeneous evacuation platforms results in complex patient evacuation dynamics that prove difficult to model in closed form. By contrast, our decision support system and accompanying model contains only two islands, idealized evacuation asset resources, and unlimited treatment facility capacity. While this simplified system sufficiently demonstrates the advantages of shore to ship to shore patient transfer, it is not suitable for coordinating evacuation operations at scale. The future development of more intricate evacuation models motivates the continued use of probabilistic search algorithms, such as Monte Carlo Tree Search, in tandem with a generative environment. Value iteration and other dynamic programming solutions that require retraining with changes in environment may not be as suitable. Additionally, while we consider only a handful of evacuation platforms in our model, those same evacuation platforms are likely to be tasked to support multiple simultaneous battles and therefore have competing priorities. Understanding evacuation at the theater level, across multiple archipelagos and atolls in space and several days or weeks in time, is critical. 

\subsection{Future Work}
Unique maritime evacuation concepts arise from scaling overwater AXPs and supporting them with emerging evacuation platforms. We discuss three concepts and their interplay: the maritime evacuation chain, the maritime transfer opportunity zone, and the dedicated overwater AXP. 

\subsubsection{Maritime Evacuation Chains}
Consider the logical extension of a single watercraft enabling shore to ship to shore patient transfers. Relevant to operational reach is the radius of action, also known as the combat range, defined as half the maximum range of a given aircraft \cite{page1973range}. If a single watercraft positioned strategically is capable of bridging the radius of action between two distant aircraft, a second watercraft daisy chained to the first may extend operational reach even further. This is akin to island hopping with one's own watercraft in place of islands. Maritime evacuation chains facilitate long-distance inter-archipelago evacuation efforts across ``blue-water'', and are therefore best suited for Naval casualty evacuation platforms. By contrast, a single overwater AXP typically enables intra-archipelago evacuation operations, and may be supported by Army medical evacuation platforms. Maritime evacuation chains require a substantial level of command and control, and situational awareness, to successfully execute. Given the logistical complexity involved with multiple watercraft and aircraft operating simultaneously, maritime evacuation chains are usually worthwhile only when fixed-wing transport is unavailable. 

\subsubsection{Maritime Transfer Opportunity Zones}

The maritime transfer opportunity zone represents the area where at least one watercraft may be co-opted as an overwater AXP between two aircraft to support a given patient transfer. There may be multiple overlapping maritime transfer opportunity zones in an area of interest across several combinations of aircraft and watercraft. Transfer opportunity zones are rarely static; they are continuously reshaped as the underpinning watercraft change locations. Zones may be differentiated by the medical capabilities available at their respective watercraft, as well as by the presence of helipads and refuel capabilities. Mapping watercraft trajectories in time and space enables medical planners to generate maritime transfer opportunity zones for evacuation in preparation for future operations. Specific aircraft and watercraft can then be coupled for evacuation operations for specific time windows. This assumes the watercraft considered are underway in support of non-medical mission requirements, and therefore cannot be rerouted to optimize evacuation planning. 

\subsubsection{Dedicated Overwater Ambulance Exchange Point}

An alternative approach to evacuation planning between islands and archipelagos is to maximize maritime transfer opportunity zone coverage by strategically positioning a dedicated aviation-capable watercraft with refuel capability at an optimal location. This ``dedicated'' overwater AXP prevents maritime transfer opportunity zone blackouts, which are windows of time where the limited numbers of nearby watercraft restrict maritime evacuation operations. The optimal location for a dedicated overwater AXP will change as combat operations progress across the battlefield, as well as in response to zone coverage provided by non-dedicated watercraft. The United States Navy's Expeditionary Medical Ship is a fast transport hospital ship with a helipad and aircraft refuel capabilities, making it an excellent choice for a dedicated overwater AXP. Selecting evacuation aircraft with sufficient range for extended overwater operations further improves performance. 

Fig. 7 visualizes a hypothetical scenario involving a dedicated aviation-capable watercraft with refuel capabilities and a non-dedicated watercraft that happens into one such transfer opportunity zone. In this scenario, a patient is transported approximately 1,600 miles from Manila, Philippines to Guam. Two HH-60M Medical Evacuation Black Hawk helicopters outfitted with half-filled crash-worthy external fuel systems (CEFS) have a radius of action of 614 miles. The Philippines and Guam were selected for this example due to the frequency of movement between the two that occurs for annual military bilateral exercises Salaknib and Balikatan. The arrows indicate the direction of patient movement. The first HH-60M stationed on the dedicated aviation-capable watercraft departs the open ocean to Manila for patient pickup, then returns to its watercraft for refuel. The first HH-60M then meets a second HH-60M, based out of Guam, at a non-dedicated, underway watercraft somewhere in the maritime transfer opportunity zone to facilitate the remainder of the transfer.

\begin{figure*}[ht!]
\centering
\includegraphics[width=16.5cm]{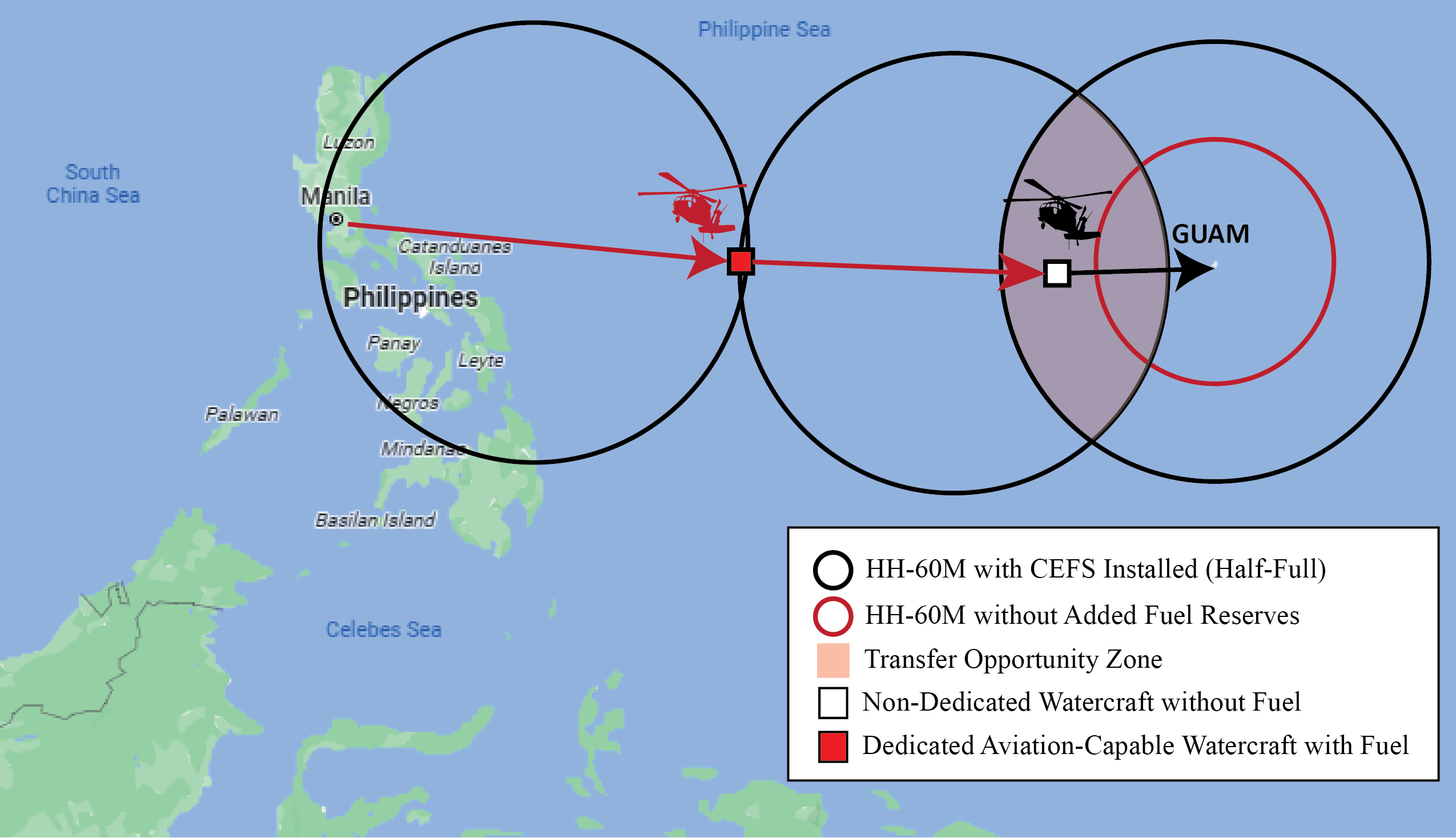}
\caption{A representative scenario showcasing the combined use of the maritime evacuation chain, maritime transfer opportunity zone, and dedicated overwater ambulance exchange point.}
\end{figure*}

\subsection{Civilian Applications}

Using watercraft to facilitate patient movement between medical aircraft resembles the larger class of cooperative heterogeneous multi-agent coordination problems, which also encompass scenarios pertinent to civilian medical evacuation. Optimizing the placement of civilian ground and air ambulance bases in support of coordinated multi-modal patient transport similarly minimizes transport times to hospitals \cite{shahriari2017bi}. Several research efforts apply heterogeneous multi-agent systems to support evacuation planning in emergency scenarios, whether by directing a fleet of aerial vehicles with different capabilities \cite{oh2018cooperative} or integrating medical and temporal data from an evolving set of sensors to support route planning \cite{taga2017multi}, \cite{tian2014use}. The strategic coordination of watercraft and medical aircraft perhaps most directly benefits large-scale evacuation planning for isolated island and coastal communities prior to or in the aftermath of natural disasters. Maritime evacuation remains substantially understudied \cite{krutein2022evacuating}, \cite{islam2023developing}, despite the increased risk of natural disasters caused by global warming \cite{ipcc2012managing}. The ensuing hazards are especially dangerous for remote communities that rely primarily on limited and vulnerable waterways or roadways \cite{krutein2022isolated}. 


\subsection{Conclusion}
Medical evacuation in a maritime theater is inherently challenging. Using watercraft as overwater AXPs to transfer patients between aircraft helps overcome the tyranny of distance and provides flexibility in operations. While promising, overwater AXPs are fundamentally different than traditional fixed land AXPs due to their underway and often non-medical nature. We partner with the United States Army's 25th Combat Aviation Brigade to conduct MEDEVAC Projects Week, a first of its kind demonstration of the overwater AXP using actual aircraft and watercraft in the Hawaiian Islands. Finally, we introduce three extensions of the overwater AXP to inspire future evacuation capabilities, and consider civilian applications.

\noindent
\textit{Notice: Mention or description of any private company or product must not be interpreted as Federal or DoD endorsement.}

\newpage

\bibliography{main}

\end{document}